\documentclass[twocolumn,aps,showpacs]{revtex4}
\usepackage{amssymb}
\usepackage{amsmath,bm}
\usepackage{graphicx}
\usepackage[normalem]{ulem}
\usepackage{multirow}

\usepackage[usenames, dvipsnames]{xcolor}
\usepackage{tensor}

\renewcommand{\sout}{\bgroup \color{red} \ULdepth=-.5ex \ULset}

\begin{document}

\title{Charmed hadron production in    an improved  quark coalescence  model} 

\author{Sungtae Cho\footnote{%
sungtae.cho@kangwon.ac.kr}}
\affiliation{Division of Science Education, Kangwon National University, Chuncheon 24341, Korea}

\author{Kai-Jia Sun\footnote{%
Corresponding author: kjsun$@$tamu.edu}}
\affiliation{Cyclotron Institute and Department of Physics and Astronomy, Texas A\&M University, College Station, Texas 77843, USA}

\author{Che Ming Ko\footnote{%
ko@comp.tamu.edu}}
\affiliation{Cyclotron Institute and Department of Physics and Astronomy, Texas A\&M University, College Station, Texas 77843, USA}

\author{Su Houng Lee\footnote{%
suhoung@yonsei.ac.kr}}
\affiliation{Department of Physics and Institute of Physics and Applied Physics, Yonsei University, Seoul 03722, Korea}

\author{Yongseok Oh\footnote{%
yohphy@knu.ac.kr}}
\affiliation{Department  of  Physics,  Kyungpook  National  University,  Daegu  41566,  Korea}
\affiliation{Asia Pacific Center for Theoretical Physics, Pohang, Gyeongbuk 37673, Korea}

\date{\today}

\begin{abstract}
We study the production of charmed hadrons $D^{0}$ and $\Lambda_c^+$ in relativistic heavy ion collisions using 
 the charm quark coalescence.  Besides taking into consideration of changing hadron sizes in hot dense medium, which results in an enhanced coalescence probability for charm quarks of very low transverse momenta,   we  also  include  the collective flow effect on heavier resonances, which leads to a shift of massive charmed resonances to larger transverse momenta. Including the conversion of charm quarks not undergoing coalescence to hadrons by independent fragmentation, we obtain a good description of  the measured yield ratio  $\Lambda_c^+/D^0$ as a function of  transverse momentum  in $\nuclide{Au} + \nuclide{Au}$ collisions at $\sqrt{s_{NN}}^{}=200$~GeV by the STAR Collaboration at the Relativistic Heavy Ion Collider.
\end{abstract}

\pacs{25.75.-q, 25.75.Dw}
\maketitle

\section{Introduction}\label{introduction}

The main goal of relativistic heavy ion collisions,  such as those being carried out at Relativistic Heavy Ion Collider (RHIC) and the Large Hadron Collider (LHC), is to explore the phase diagram of    matter described by  quantum chromodynamics, especially the properties of deconfined quark-gluon plasma (QGP) that could be created in these collisions, and its transition to hadronic matter~\cite{Jacak:2012dx,Shuryak:2014zxa}.    Although the bulk properties of the created QGP are governed by light quarks and gluons, the rare heavy charm and bottom quarks produced in ultra-relativistic heavy ion collisions are also useful probes of its properties~\cite{Linnyk:2008hp,He:2012df,He:2011qa,Uphoff:2012gb,Cao:2015hia,Nahrgang:2014vza,Scardina:2017ipo,Das:2017dsh,Das:2016cwd,Das:2015ana,Das:2013kea,Tolos:2016slr,Cao:2016gvr,Beraudo:2017gxw,Beraudo:2014boa,Alberico:2011zy}.   Studying resulting charmed and bottom hadrons, such as the $D$ ($B$) mesons and $\Lambda_{c}$ ($\Lambda_{b}$), $\Sigma_{c}$ ($\Sigma_{b}$), $\Xi_{c}$ ($\Xi_{b}$) baryons in relativistic heavy ion collisions, has thus been a topic of great interest~\cite{Plumari:2017ntm,Ravagli:2007xx,He:2019tik}. In recent experiments by the STAR Collaboration, the transverse moment spectrum of $D^0$ mesons and also the $\Lambda_c/D^0$ ratio from $\nuclide{Au} + \nuclide{Au}$ collisions have been measured~\cite{Dong:2017dws,Xie:2017jcq,Zhou:2017ikn,Xie:2018thr}.   The  experimental data from collisions at $10$-$80$\% centrality shows  the ratio  $\Lambda_c^+/D^0 \simeq 0.8$--$1.1$  in the transverse momentum region of  $3 < p_{T}^{} < 6$~GeV, which is a very large enhancement compared to the value predicted  from the fragmentation  of charm quarks or from the PYTHIA results for $p+p$ collisions~\cite{Lisovyi:2015uqa,Sjostrand:2006za}.  Such a ratio is also much larger than the prediction for the integrated yield from the  statistical hadronization model, where $\Lambda_{c}^+/D^{0} \, \simeq 0.25 - 0.3$~\cite{Kuznetsova:2006bh,Andronic:2007zu,Andronic:2010dt}.

Similar enhancements of the baryon to meson ratios of hadrons consisting of light and strange quarks in relativistic heavy ion collisions compared to those from $p+p$ collisions were previously seen in experiments at 
RHIC~\cite{Agakishiev:2011dc,Agakishiev:2011ar,Abelev:2013xaa,Abelev:2014laa}, and they were successfully explained in terms of the quark coalescence model for the production of hadrons of intermediate momenta~\cite{Greco:2003xt,Fries:2003vb,Greco:2003mm,Fries:2008hs,Minissale:2015zwa}. Extending the quark coalescence model to charm quarks, it was shown in Refs.~\cite{Lee:2007wr,Oh:2009zj} that the $\Lambda_{c}^+/D^0$ ratio in relativistic heavy ion collisions is also enhanced when compared with that in $p+p$ collisions at the same energy.  An improved study using a more realistic charm quark  spectrum was later carried out in Ref.~\cite{Plumari:2017ntm}. The predicted ratio $\Lambda_{c}^+/D^0$ at $p_T^{} \approx 6$~GeV from this study is found to be about 0.4, which is still a factor of 2 smaller than the measured value in  the  STAR experiments. Recently, it was found that this ratio could be explained by the resonance recombination model (RRM) ~\cite{Ravagli:2007xx}
%~\cite{Ravagli:2007xx,He:2019vgs} 
after including a large number of “missing” charm baryon states ~\cite{He:2019vgs}. 

In the present study, we improve   the work of Ref.~\cite{Oh:2009zj} by employing a more realistic charm quark spectrum and also including in the quark coalescence model the flow effect  on produced heavy particles.  In the usual coalescence model, such as the one employed in Refs.~\cite{Oh:2009zj,Plumari:2017ntm}, the transverse momentum of a produced hadron is  equal to the total momentum of coalesced quarks. As a result, hadrons of different masses formed from these quarks all have same momentum, which is in contrast  to the hydrodynamical picture that hadrons of larger masses are shifted to higher transverse momentum as a result of collective flow. To include this effect, we boost a produced hadron from the center of mass of coalescing quarks, where its Wigner function is calculated to give its formation probability, back to the fireball frame using the physical mass of the hadron.  In this way, the momenta of produced hadrons, particularly resonances of large masses, are increased by the effect of parton collective flow.  With this improved approach as well as after including possible increase of hadron sizes in hot dense medium and the fragmentation contribution from charmed quarks not used in coalescence, we obtain a good description of the measured $D^0$ momentum spectrum and the predicted $\Lambda_{c}^+/D^0$ ratio as a function of $p_T^{}$   also agrees nicely with the available data    from  RHIC without assuming the existence of missing high mass  charmed baryon resonances as in Ref.~\cite{He:2019vgs}.  In addition, we find that the total yield ratio $\Lambda_{c}^+/D^0$ is around 0.64, and the ratio $\Lambda_{c}^+/D^0$ at $p_T^{}=8$~GeV can be as large as 0.6, which is much larger than the predictions from   previous studies reported in Refs.~\cite{Oh:2009zj,Plumari:2017ntm}.

\section{Quark momentum spectra}

\subsection{Light quarks  }

For the light quark momentum spectra, we adopt an approach  similar to that employed in Ref.~\cite{Oh:2009zj} by using  more realistic  ones from Ref.~\cite{Plumari:2017ntm}.  Specifically, the longitudinal momentum distribution of light quarks is assumed to be boost invariant in the rapidity range of $|y|\le 0.5$.  To take into account the collective flow of quark-gluon plasma, we assume  that  light  partons have a radial flow profile of $\beta_T(r_T^{})=\beta_{\rm max} {r_T^{}}/{R}$ in the transverse plane of a heavy ion collision, where $R$ is the transverse radius of the quark-gluon plasma at hadronization proper time $\tau$.  The transverse momentum distribution of light quarks is taken to be a thermal one  at temperature $T=165$~MeV, that is 
\begin{eqnarray}
\label{quark-distr}
\frac{dN_{q,\bar{q}}}{\:d^{2}p_{T}^{}} &=& \frac{g_{q,\bar{q}} \, \tau m_{T}^{}}{(2\pi)^{3}} \nonumber\\
&\times&\int \exp \left[-\frac{\gamma_{T}^{}(m_{T}^{}-p_{T}^{} \cdot \beta_{T}^{})\pm\mu}{T} \right] d^{2}r_{T}^{}. \label{light-quark-dis}
\end{eqnarray}
In the above, $g_{q,\bar{q}}=6$ are the spin-color degeneracies of  quarks and antiquarks, $\mu$   is the quark baryon chemical potential with the plus and minus signs for quarks and antiquarks, respectively, $m_T^{} = \sqrt{p_T^2+m_{q,\bar{q}}^2}$ is the transverse mass with $m_{q,\bar q}$ being the constituent quark and antiquark masses, which are taken to be 300 MeV and 475 MeV for light and strange quarks, respectively, and $\gamma_T=1/\sqrt{1-\beta_T^2}$.   As in Refs.~\cite{Greco:2003xt,Minissale:2015zwa}, we also include the contribution from gluons in the quark-gluon plasma, which are taken to have a   distribution similar to that of light quarks, by converting them to quarks and anti-quarks according to the quark flavor compositions in the quark-gluon plasma. The parameters for describing the fireball of hadronizing quark-gluon plasma and the numbers of light and strange quarks and antiquarks are summarized in Table~\ref{tabQuark},  and their values are very similar to those used in Ref.~\cite{Plumari:2017ntm}. 

\begin{table} [t]
\begin{center}
\begin{tabular}{lc|c|c|c|c|c|c} 
\hline \hline
%          parameters  & $N_u$ ($N_{\bar{u}}$) & $N_s$ &   $V$ (fm$^3$)  & $T$ (MeV)  & $\beta_{\rm max}$   \\ 
& $N_u$ ($N_{\bar{u}}$) & $N_s$ &   $R$ (fm)  & $\tau$ (fm/$c$) & $T$ (MeV)  & $\beta_{\rm max}$   & $\mu$ (MeV) \\ 
\hline          
 & 243 (224) & 143  & 8.5 & 4.5 &   165            &  0.5  & 10    \\ 
\hline \hline
\end{tabular}
\end{center}
\caption{\protect Up and antiup quark numbers $N_u$ and $N_{\bar u}$, strange quark number $N_s$,  transverse radius $R$, hadronization proper time $\tau$, temperature $T$, and flow velocity $\beta_{\rm max}$ of QGP produced at mid-rapidity of central $\nuclide{Au} + \nuclide{Au}$ collisions at $\sqrt{s_{NN}} = 200$~GeV and 0-10\% centrality.  The number of down and antidown quarks are the same as those of up and antiup quarks.} 
\label{tabQuark}
\end{table}
 
\subsection{Charm quarks  }

\begin{table} [b]
\begin{center}
\begin{tabular}{l|c|c|c|c|c|cc} 
\hline \hline
          RHIC  & $a_0^{}$ & $a_1^{}$ & $a_2^{}$ & $a_3^{}$  & $a_4^{}$  & $a_5^{}$   \\ 
\hline
 $p_T^{} \leq p_0^{}$  & 0.69 & 1.15  & 1.57  &   ---     &   ---     &   ---     \\ 
 $p_T^{} \geq p_0^{}$  & 1.08 & 3.04  & 0.71  & 9.914   & 2.5   & 3.48  \\
\hline \hline
\end{tabular}
\end{center}
\caption{Parameters  used in the parametrization of charm quark transverse momentum spectrum at mid-rapidity   
of central $\nuclide{Au} + \nuclide{Au}$ collisions at $\sqrt{s_{NN}^{}} = 200$~GeV.}
\label{tabCHARM}
\end{table}

For the charm quark momentum spectrum in heavy ion collisions at RHIC, we use the one parametrized in Ref.~\cite{Plumari:2017ntm}, which is based on   results from a transport model study of charm quark energy loss and flow.  It has the form  
\begin{eqnarray}\label{charm}
&&\frac{dN_c}{d^2p_T^{}}\nonumber\\
&&=\left\{\begin{array}{lr}
a_0^{} \exp{[-a_1^{} p_T^{a_2^{}}]}, & p_T^{} \leq p_0^{} \\
a_0^{} \exp{[-a_1^{} p_T^{a_2^{}}]}+a_3^{} \left(1+p_T^{a_4^{}} \right)^{-a_5^{}}, & p_T^{} > p_0^{} 
\end{array}\right. 
\nonumber \\
\end{eqnarray}
where $p_0^{} = 1.85$~GeV and the values of the parameters  $a_i$ with $i=1,\cdots, 5$ are given in Table~\ref{tabCHARM}.  They are slightly different from those in Ref.~\cite{Plumari:2017ntm} to achieve a better description of the measured $D^0$ spectrum at large transverse momentum.  Integrating the above transverse momentum spectrum gives the total number of heavy quarks of $dN_c/dy \simeq 2.1$   for the collisions at  RHIC considered in the present study.   For the charm quark mass, we use $m_c = 1.5~\mbox{GeV}$ in the present study.

\section{Quark coalescence}
For simplicity, we assume as in Ref.~\cite{Oh:2009zj} that the spatial distribution  of quarks   is uniform  in the thermalized QGP inside a fire cylinder of volume $V=\pi R^2\tau$.  Taking the Wigner function of hadrons to be Gaussian in space and in momentum and neglecting the space and velocity correlation of light quarks due to collective flow, we can  integrate out the spatial part  of the coalescence formula   and obtain the transverse momentum spectrum of produced heavy mesons of certain species as
\begin{eqnarray}
\frac{dN_M}{d\bm{p}_M^{}} &=& g_M^{} \frac{(2\sqrt{\pi}\sigma)^3}{V}
\int d \bm{p}_1^{} d \bm{p}_{2}^{} \frac{dN_1}{d\bm{p}_1^{}}
\frac{dN_{2}}{d\bm{p}_{2}^{}} 
\nonumber \\ && \mbox{} \times
\exp\left(-\bm{k}^2 \sigma^2 \right) \delta({\bf p}_M-{\bf p}_1-{\bf p}_2). \label{eq:meson-coal}
\end{eqnarray}
In the above, $g_M^{}$ is the statistical factor for colored spin-1/2 quark and antiquark to form a color neutral meson, e.g., $g_{D^0}^{}=1/36$ and $g_{D^{*0}}^{}=1/12$ for $D^0$ and $D^{*0}$, respectively. The momenta ${\bf p}_1$, ${\bf p}_2$, and ${\bf p}_M$ are those of the heavy quark, light quark, and produced heavy meson, respectively, with the $\delta$ function to ensure the momentum conservation. The relative transverse momentum $\bm{k}$ between the heavy quark of mass $m_1$  and light antiquark of mass $m_2$  is defined as
\begin{equation}
\bm{k} = \frac{1}{m_1^{} + m_2^{}}
\left( m_2^{} \bm{p}_1' - m_1^{} \bm{p}_2' \right),
\label{eq:def_k}
\end{equation}
where ${\bf p}_1^\prime$ and ${\bf p}_2^\prime$ are the momenta of the heavy quark and light antiquark in the center of mass frame of produced heavy meson.   
The width parameter $\sigma$ is related to the harmonic oscillator frequency $\omega_M$ by $\sigma= 1/\sqrt{\mu\omega_M}$ with $\mu = m_1^{} m_2^{} / (m_1^{} + m_2^{})$ being the reduced mass.

%The value of $\sigma$ can be related to the size of produced hadron and will be given in the next section.

Similarly, the  momentum spectrum of heavy baryons from the coalescence of a charm quark and two light quarks can be calculated according to 
\begin{eqnarray}
\frac{dN_B}{d\bm{p}_{B}^{}} &=& g_{B}^{} \frac{(2\sqrt{\pi})^6
(\sigma_1^{} \sigma_2^{})^3}{V^2} \int d \bm{p}_1^{} d \bm{p}_2^{} d \bm{p}_3 
\frac{dN_1}{d\bm{p}_1^{}} \frac{dN_2}{d\bm{p}_2^{}} \frac{dN_3}{d\bm{p}_3^{}} 
\nonumber \\ && \mbox{} \times 
\exp\left(-\bm{k}_1^2 \sigma_1^2 - \bm{k}_2^2 \sigma_2^2 \right) \delta({\bf p}_B-{\bf p}_1-{\bf p}_2-{\bf p}_3),
\nonumber \\
\label{eq:3q-coal}
\end{eqnarray}
where the index $3$ refers to the heavy quark and indices $1$ and $2$ refer to light quarks, and $g_B^{}$ is the statistical factor, which, for example, is $1/108$ for $\Lambda_c$, $1/36$ for $\Sigma_c$, $1/54$ for $\Xi^c (\Xi_c^\prime)$, and $1/8$ for $\Sigma_c^*$ and $\Xi_c^*$.  The relative transverse momenta are defined as
\begin{eqnarray}
\bm{k}_1 &=& \frac{1}{m_1^{} + m_2^{}}
\left( m_2^{} \bm{p}_1' - m_1^{} \bm{p}_2' \right),
\nonumber \\
\bm{k}_2 &=& \frac{1}{m_1^{} + m_2^{} + m_3^{}}
\left[ m_3^{} \left( \bm{p}_1' + \bm{p}_2' \right)
- (m_1^{} + m_2^{}) \bm{p}_3' \right],
\nonumber \\
\end{eqnarray}
with ${\bf p}_1^\prime$, ${\bf p}_2^\prime$ and ${\bf p}_3$ being the momenta of the heavy quark and two light quarks in the center of mass frame of produced heavy baryon.
The width parameters $\sigma_i^{}$ are related to the oscillator parameter $\omega_B$ by $\sigma_i^{} =1/\sqrt{\mu_i\omega_B}$ with
\begin{eqnarray}
\mu_1^{} = \frac{m_1^{} m_2^{}}{m_1^{} + m_2^{}}, \qquad
\mu_2^{} = \frac{(m_1^{} + m_2^{}) m_3^{}}{m_1^{} + m_2^{} +m_3^{}}.
\end{eqnarray}

As in Ref.~\cite{Oh:2009zj}, we take the oscillator constants  $\omega_M$ for  $D_0$ meson and $\omega_B$ for  $\Lambda_c^+$ baryon as parameters, and determine their values  by fitting the spectrum of $D_0$ meson and requiring all the charm quarks at low momenta to hadronize by quark coalescence. To include the flow effect on produced hadrons, we carry out the coalescence calculation  at the medium rest frame and then boosting these hadrons to the fireball frame. Because of the smaller quark thermal velocity than the flow velocity, the flow  effect can be approximately included by first calculating the formation probability of a charmed hadron from coalescing charm and light quarks using its Wigner function evaluated in the center of mass of these quarks and then boosting the resulting charm hadron to the fireball frame using its physical mass. This approximation results in the multiplication of the momentum ${\bf p}_M$ and ${\bf p}_B$ in the delta functions in Eqs.(3) and (4) by the factor  $(E_1+E_2)/E_M$ and $(E_1+E_2+E_3)/E_B$, respectively, where $E_i=\sqrt{m_i^2+{\bf p}_i^2}$ and $E_{M,B}=\sqrt{m_{M,B}}^2+{\bf p}_{M,B}^2$ with $m_{M,B}$ being the mass of produced heavy hadrons. In this case, heavy resonances with large masses would have large transverse momenta in the rest frame of the expanding QGP, which is consistent with the hydrodynamic picture  that  the additional momenta acquired by particles due to the collective flow are larger if they are more massive.  This effect has been neglected in previous studies
%~\cite{Oh:2009zj,Plumari:2017ntm} 
based on the coalescence approach~\cite{Oh:2009zj,Plumari:2017ntm} where the transverse momentum spectrum of produced particles is independent of their masses. The present approach is thus more appropriate for studying the production of massive resonances in relativistic heavy ion collisions. We note that the production of massive hadrons is not suppressed in the coalescence model as it is based on the sudden approximation. This is in contrast to that in the resonance recombination model of Ref.~\cite{Ravagli:2007xx} used in Ref.~\cite{He:2019vgs} due to the required energy conservation in this approach. 

\section{charm quark fragmentation}

Similarly to Refs.~\cite{Oh:2009zj,Plumari:2017ntm}, charm quarks that are not used for producing hadrons via  coalescence with light quarks are  converted to hadrons by fragmentation. In terms of the fragmentation probability $P_{\rm frag}(p_T^{})=1-P_{\rm coal}(p_T^{})$ of a charm quark of transverse momentum $p_T^{}$, where $P_{\rm coal}(p_T^{})$ is its probability to coalesce with light quarks, the momentum spectrum of certain hadron species from the fragmentation of non-coalesced charm quarks is given by
\begin{equation}
\frac{dN_{\rm had}}{d^{2}p_T^{}}=\sum \int dz P_{\rm frag}(P_T^{})\frac{dN_{N_c}}{d^{2}p_T^{}} 
\frac{D_{{\rm had}/c}(z,Q^{2})}{z^{2}}.
\label{Eq:frag}
\end{equation}
In the above,  $z=p_{\rm had}/p_{c}$ is the fraction of charm quark momentum carried by the produced hadron and $Q^2=(p_{\rm had}/2z)^2$ is the momentum scale for the fragmentation process. For  the fragmentation function $D_{\rm had/c}(z,Q^{2})$,  we use the one from Ref.~\cite{PSSZ83},
\begin{eqnarray}
D_{\rm had}(z) \propto 1/\left[ z \left(1-\frac{1}{z}-\frac{\epsilon_c}{1-z} \right)^2 \right],
\end{eqnarray}
with $\epsilon_c$  being a free parameter to fix the shape of the fragmentation function. In the present  study, we choose $\epsilon_c=0.006$ for $D$ mesons and $\epsilon_c=0.02$ for $\Lambda_c$ baryons, which leads to the fragmentation branching ratios to $D^0$, $D^+$, $D_s^+$, and $\Lambda_c^+$ being 0.607, 0.196, 0.121, and 0.076, respectively~\cite{Oh:2009zj}. 

\section{results}

\begin{table} [t]
\begin{center}
\begin{tabular}{c|c|c|c|cc}
\hline
Meson & Mass (MeV) & $I (J)$ & \\
\hline 
$D^+ =\bar{d}c$		& 1869 & $\frac{1}{2} \,(0)$	&\\
$D^0 =\bar{u}c$ 	& 1865 & $\frac{1}{2} \,(0)$	&\\
$D_{s}^{+} =\bar{s}c$	& 1968 & $0 \,(0)$		& \\
\hline
Resonances & & & Decay modes& B.R. \\
\hline
$D^{*+} =\bar{d}c$	& 2010 & $\frac{1}{2} \, (1)$	& $D^0 \pi^+$ & $68\%$ \\
&&& $D^+ X$& $32\%$ \\
$D^{*0} =\bar{u}c$	& 2007 & $\frac{1}{2} \, (1)$	& $D^0 \pi^0$ & $62\%$ \\
&&& $D^0 \gamma$& $38\%$ \\
$D_{s}^{*+} =\bar{s}c$	& 2112 & $0 \, (1)$		& $D_{s}^+ X$ & $100\%$ \\
\hline
\end{tabular}
\end{center}
\caption{
Charmed mesons considered in the present study. The branching ratios (B.R.) of resonances decaying to the ground states are taken from Ref.~\cite{Agashe:2014kda}.
\label{tab:D}}
\end{table}

\begin{table} [t]
\begin{center}
\begin{tabular}{c|c|c|c|cc}
\hline
Baryon & Mass (MeV) & $I (J)$ & \\
\hline 
$\Lambda_c^+ =udc$	& 2286 & $0 \, (\frac{1}{2})$	&\\
$\Xi_c^+ =usc$	& 2467 & $\frac{1}{2} \, (\frac{1}{2})$	&\\
$\Xi_c^0 =dsc$	& 2470 & $\frac{1}{2} \, (\frac{1}{2})$	&\\
\hline
Resonances & & & Decay modes& B.R. \\
\hline
$\Sigma_c^0 =ddc$	& 2455 & $1 \, (\frac{1}{2})$	&$\Lambda_c^+ \pi^-$ & $100\%$\\
$\Sigma_c^+ =udc$	& 2455 & $1 \, (\frac{1}{2})$	&$\Lambda_c^+ \pi^0$ & $100\%$\\
$\Sigma_c^{++} =uuc$	& 2455 & $1 \, (\frac{1}{2})$	&$\Lambda_c^+ \pi^+$ & $100\%$\\
$\Sigma_c^{*0} =ddc$	& 2520 & $1 \, (\frac{3}{2})$	&$\Lambda_c^+ \pi^-$ & $100\%$\\
$\Sigma_c^{*+} =udc$	& 2520 & $1 \, (\frac{3}{2})$	&$\Lambda_c^+ \pi^0$ & $100\%$\\
$\Sigma_c^{*++} =uuc$	& 2520 & $1 \, (\frac{3}{2})$	&$\Lambda_c^+ \pi^+$ & $100\%$\\
$\Xi_c^{*+} =usc$	& 2645 & $\frac{1}{2} \, (\frac{3}{2})$	&$\Xi_c^+ \pi^0$ & seen\\
$\Xi_c^{*0} =dsc$	& 2645 & $\frac{1}{2} \, (\frac{3}{2})$	&$\Xi_c^+ \pi^-$ & seen\\
$\Xi_c'^+ =usc$	& 2580 & $\frac{1}{2} \, (\frac{1}{2})$	&$\Xi_c^+ \gamma$ & seen \\
$\Xi_c'^0 =dsc$	& 2580 & $\frac{1}{2} \, (\frac{1}{2})$	&$\Xi_c^0 \gamma$ & seen \\
\hline
\end{tabular}
\end{center}
\caption{ Same as Table~\ref{tab:D}   but for charmed baryons.
\label{tab:Lambda}}
\end{table}

As shown in Ref.~\cite{Oh:2009zj}, the contributions from resonances to the yield of ground state hadrons are important and  should be taken into account. Tables~\ref{tab:D} and \ref{tab:Lambda} summarize  the  charmed hadrons considered in the present study, which include the ground states and the resonance states of $D$ mesons and $\Lambda_c$ and $\Xi_c$ baryons, as given by the Particle Data Group~\cite{Agashe:2014kda}. For the branching ratios  of $\Xi_c^*$ and $\Xi_c'$ baryons  decaying by strong or electromagnetic interactions to the $\Xi_c$ baryon, which are not given in Ref.~\cite{Agashe:2014kda}, they are assumed to   be 100\%.  We note that all charmed meson and baryon resonances in Tables~\ref{tab:D} and \ref{tab:Lambda} have their orbital wave functions in the $L=0$ states. 

\begin{figure}[h]
\centering
\includegraphics[scale=0.37, angle=0,clip]{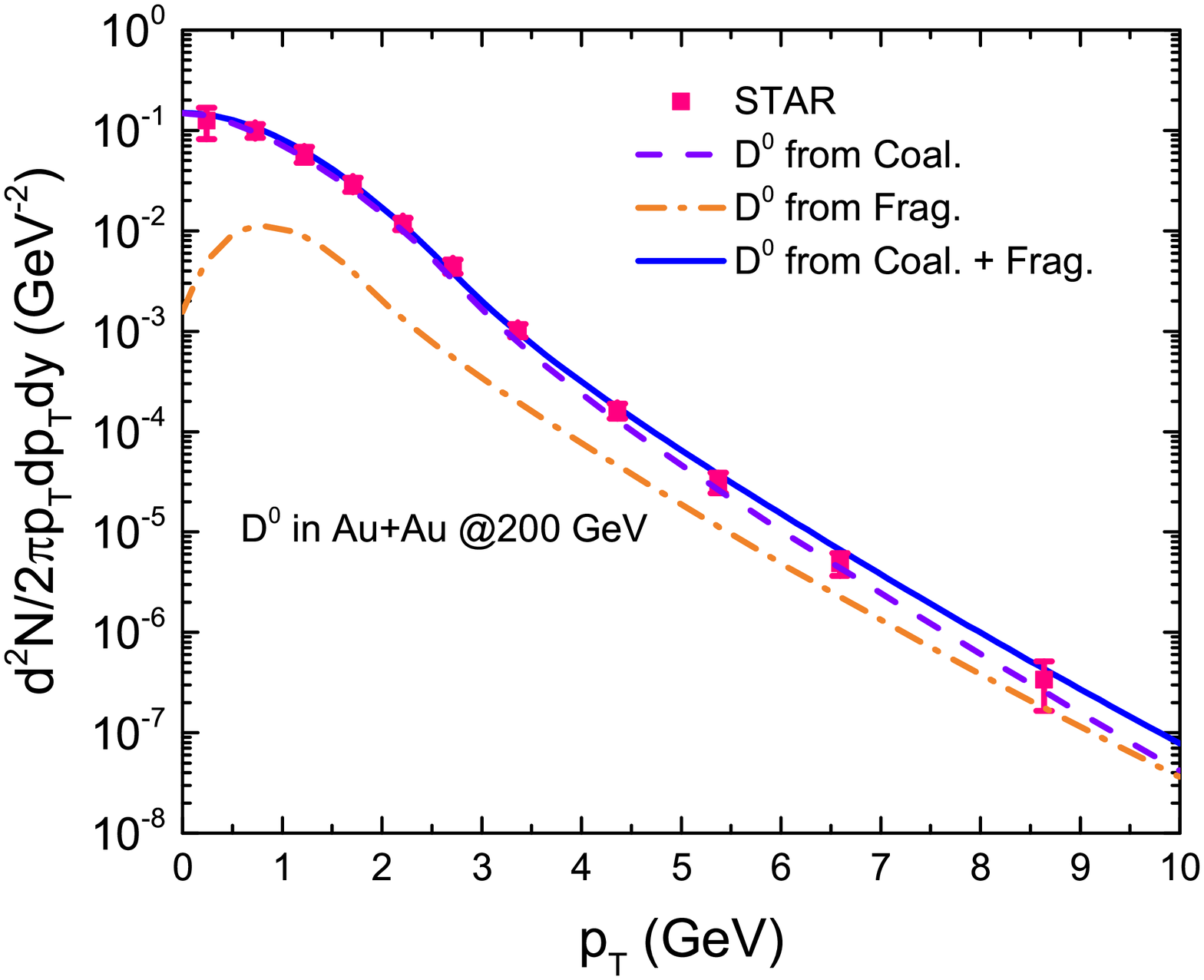}
\caption{\label{Fig1} 
Transverse momentum spectrum  of $D^0$ mesons at mid-rapidity from $\nuclide{Au} + \nuclide{Au}$ collisions at $\sqrt{s_{NN}^{}}=200$~GeV and (0--10\%) centrality. Dashed and dash-dotted lines    are  the $D^0$ spectra from charm quark coalescence and fragmentation, respectively, and their sum is given by the solid line. The experimental data shown by solid squares are taken from Ref.~\cite{Adam:2018inb}. }
\end{figure}

As in  Ref.~\cite{Oh:2009zj}, we first determine the harmonic oscillator frequency $\omega_M$ in the Wigner functions for $D^0$. The value of $\omega_M \approx 0.096$~GeV is  obtained from fitting the $D^0$ transverse momentum spectrum to the data measured by  the  STAR Collaboration. This value is about a factor of 3 smaller than the value 0.33 GeV determined from the root-mean-square charge radius 0.43 fm of $D^+$ as predicted by the light-front quark model~\cite{Hwang:2001th}, implying an increase of the $D^0$ charge radius by a factor of 1.85 at temperature of $165$ MeV. We note that our oscillator constant for the charmed meson is close to the value of 0.106 GeV used in Ref.~\cite{Oh:2009zj}.  Figure~\ref{Fig1} shows our results for the $D_0$ transverse momentum spectrum from  charm  quark coalescence (dashed line), fragmentation (dash-dotted line), and their sum (solid line). It shows that the contribution from  charm quark coalescence dominates at $p_T^{} <  10$~GeV while that of fragmentation takes over at $p_T^{}$ larger than around 10~GeV.  This behavior is quite different from that obtained  in Ref.~\cite{Plumari:2017ntm}  with the conventional coalescence model,  which neglects the effect of flow on the momenta of produced charmed hadrons and thus results in a significant contribution from   charm quark fragmentation already at $p_T^{}> 3$~GeV.

\begin{figure}[h]
\centering
\includegraphics[scale=0.36,angle=0,clip]{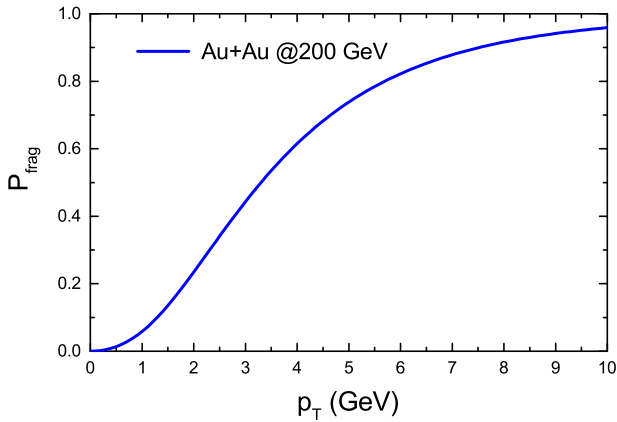}
\caption{\label{Fig2}
Fragmentation probability of charm quarks as a function of transverse momentum for central Au+Au collisions at $\sqrt{s_{NN}}=200 \, \mbox{GeV}$ and  0-10\% centrality.}
\end{figure}

Figure~\ref{Fig2} shows the fragmentation probability $P_{\rm frag}$ of charm quarks as a function of transverse momentum in central $\nuclide{Au} + \nuclide{Au}$ collisions at $\sqrt{s_{NN}^{}}=200$~GeV.  Although charm quarks with $p_T^{} > 4$~GeV are more likely to hadronize by fragmentation than coalescence, only $D_0$ mesons of $p_T^{} > 10$~GeV are mainly produced by charm quark fragmentation as shown in Fig.~\ref{Fig1}.  This is because the charm quark in $D^0$ from coalescence (fragmentation) mainly comes from those with momentum smaller (larger) than that of $D^0$.

\begin{figure}[h]
\centering
\includegraphics[scale=0.35, angle=0,clip]{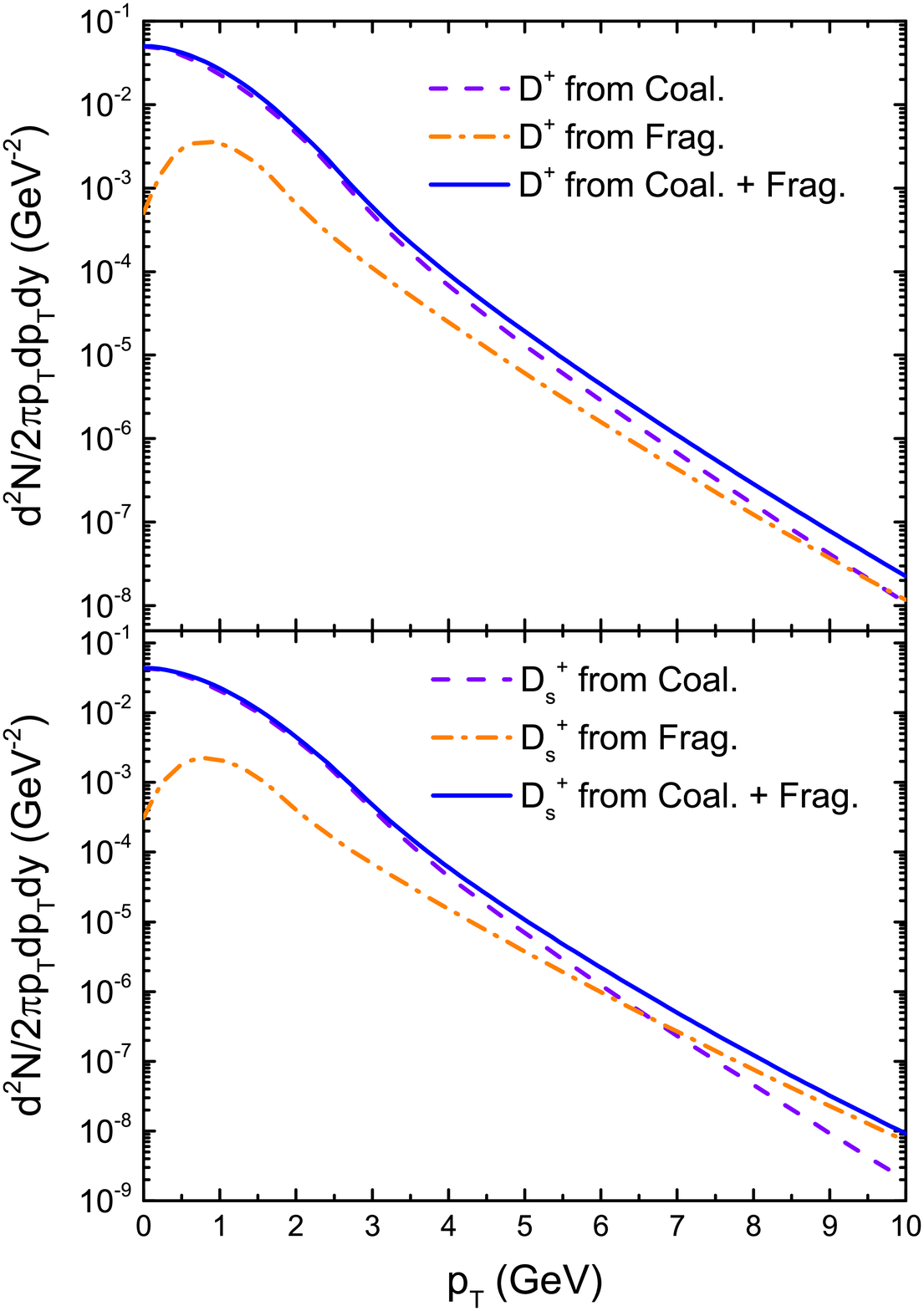}
\caption{\label{Fig3} 
\protect Transverse momentum spectra of $D^+$ \protect  (upper panel) and $D_s^+$ (lower panel) mesons in central Au+Au collisions at $\sqrt{s_{NN}^{}}=200$~GeV and  0-10\% centrality.  Dashed and dash-dotted lines   are  the spectra from charm quark coalescence and fragmentation, respectively, and their sum is shown by the solid line.}
\end{figure}

\begin{table} [b]
\begin{center}
\begin{tabular}{l|c|c|c|c|cc} 
\hline \hline
          yield  & ${D^0}$ & ${D^+}$ &   ${D_s^+}$   & $\Lambda_c^+$  & $\Xi_c$ \\ 
 \hline         
 RHIC  &0.85 &0.275  &0.236          &0.547  &0.175   \\ 
\hline \hline
\end{tabular}
\end{center}
\caption{Charmed hadron yields in central Au+Au collisions at $\sqrt{s_{NN}}=200 \, \mbox{GeV}$ and  0-10\% centrality.}
\label{RHICYield}
\end{table}

We also compute the spectra of produced $D_s^+$ and $D^+$ mesons, and the results are presented in Fig.~\ref{Fig3}. It is found that although the fragmentation contribution dominates at $p_T^{} > 10$~GeV for $D^+$, which is similar to that for $D_0$ shown in Fig.~\ref{Fig1}, it becomes important already at $p_T^{} > 7$~GeV for $D_s^+$. This is due to the softer $D_s$ transverse momentum spectrum than that of $D_0$ from  charm quark coalescence, which is also seen in Ref.~\cite{Zhao:2018jlw}. The yields of various charmed hadrons are summarized in Table~\ref{RHICYield}, which shows that the total number of charmed mesons is about 1.36 with the number of $D_0$ about three times that of $D^+$ because of the dominant contribution from the decay of charmed meson resonances.  For the remaining 0.74 charm quarks, they are converted to charmed baryons by coalescence and fragmentation as described below.  

\begin{figure}[h]
\centering
\includegraphics[scale=0.36, angle=0,clip]{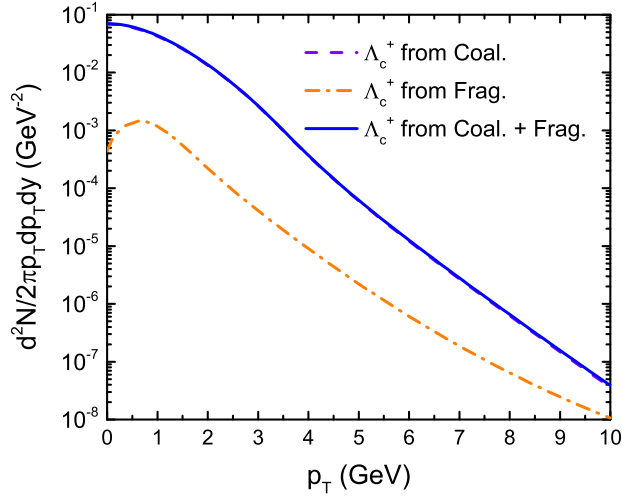}
\caption{\label{Fig4} 
\protect Transverse momentum spectra of $\Lambda_c^+$ baryon in Au+Au collisions at $\sqrt{s_{NN}}=200 \, \mbox{GeV}$ and  0-10\% centrality. Dashed and dash-dotted lines  are  the $\Lambda_c^+$ spectra  from charm quark coalescence and fragmentation, respectively, and the solid line is their sum.}
\end{figure}

Shown in Fig.~\ref{Fig4} is the $\Lambda_c^+$ spectrum, which includes those  from coalescence (dashed line), fragmentation (dash-dotted line), and their sum (solid line).  These results are obtained with the oscillator parameter $\omega_B = 0.16$~GeV for charmed baryons to ensure that the remaining charm quarks of very low transverse momenta, which are not used in the production of charmed mesons from the coalescence of charm quark with light antiquarks, are all used in the production of charmed baryons. This value of $\omega_B$ corresponds to an increase of the sizes of charmed baryons at temperature of 165 MeV compared to their values in free space using $\omega_B=0.33$.  This then leads to a yield of 0.547 for $\Lambda_c^+$ and 0.175 for $\Xi_c$ as shown in Table V.  The total integrated yield ratio $\Lambda_c^+/D_0$ is then about 0.64, which is slightly larger than the value of about 0.54 measured in $p+p$ collisions at $\sqrt{s}= 7$~TeV at LHC~\cite{Acharya:2017kfy,He:2019tik}. We note that the oscillator constant used here for charmed baryons is larger than that for charmed mesons, which is different from that in Ref.~\cite{{Oh:2009zj}} where they are taken to have the same value.

\begin{figure}[h]
\centering
\includegraphics[scale=0.36, angle=0,clip]{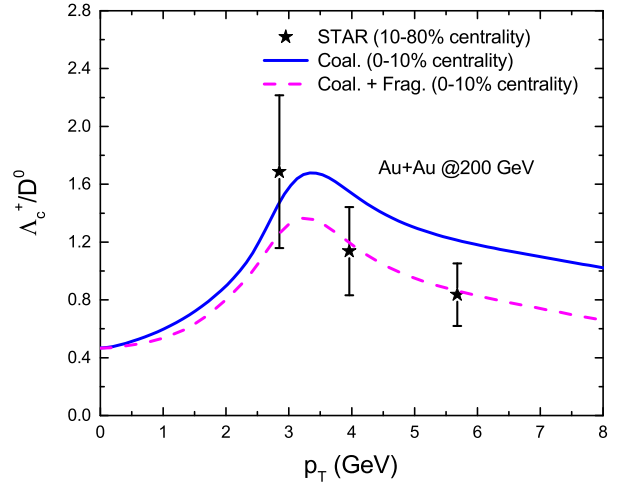}
\caption{\label{Fig5} 
The yield ratio $\Lambda_c^+/D^0$ as a function of transverse momentum for $\nuclide{Au} + \nuclide{Au}$ collisions at $\sqrt{s_{NN}^{}}=200$~GeV. Solid and dashed lines    denote  the ratio from only  charm quark coalescence and the sum of charm quark  fragmentation and coalescence   contributions for collisions at 0-10\% centrality.  The experimental data from Ref.~\cite{Xie:2018thr} for the 10-80\% centrality are shown by solid stars with combined statistical and systematic uncertainties.}
\end{figure}

In Fig.~\ref{Fig5}, we show the yield ratio $\Lambda_c^+/D^0$ as a function of transverse momentum in $\nuclide{Au} + \nuclide{Au}$ collisions at $\sqrt{s_{NN}^{}}=200$~GeV and  0-10\% centrality. It is seen that the fragmentation contribution suppresses this ratio, and the results from the sum of coalescence and fragmentation contributions (dashed line) describe very well the experimental data of Ref.~\cite{Xie:2018thr}.  For $p_T^{} \approx 6$~GeV, the ratio $\Lambda_{c}^+/D^0$ is predicted to be close to 1.0, which is much higher than the value of less than 0.4 obtained in Refs.~\cite{Oh:2009zj,Plumari:2017ntm}.  In particular, we find that the ratio $\Lambda_{c}^+/D^0$ at $p_T^{} = 8$~GeV can be as large as 0.6, while it was predicted to be around $0.2$ in Ref.~\cite{Plumari:2017ntm}.

Compared to the previous studies reported in Refs.~\cite{Oh:2009zj,Plumari:2017ntm}, the contribution from charm quark fragmentation in  the present study  is less important   due to the inclusion of the flow effect  on the momenta of hadrons formed from quark coalescence, which shifts higher mass charmed baryon resonances to larger transverse momenta, and this helps describe the ratio $\Lambda_{c}^+/D^0$  in the transverse momentum region of $4<p_T<6$ GeV.  

\section{Conclusions}

Using the charm quark coalescence and fragmentation model with the inclusion of the effect of collective flow on the transverse momentum spectra of produced charmed hadrons, we have studied the transverse momentum spectra of charmed mesons and baryons as well as the $\Lambda_c/D_0$ ratio.  By tuning the oscillator constants in the charmed hadron Wigner functions in the quark coalescence model, which models their changing sizes in hot dense matter, to use up all the charm quarks at $p_T^{} \approx 0$~GeV and fragmenting  the remaining charm quarks into charmed hadrons, we have obtained the ratio $\Lambda_{c}^+/D^0$ as a function of $p_T^{}$ that successfully describes the experimental data measured at RHIC. This is in contrast  to previous studies that did not include the effect of collective flow on charmed hadrons formed from quark coalescence, which underestimate substantially  this ratio at $p_T^{} > 4.5$~GeV.  Compared to results from these studies, the contribution from fragmentation is less important in the present approach. As a result, we have obtained  a much larger value for  $\Lambda_{c}^+/D^0$ at $p_T^{} > 6$~GeV than that from the conventional approach.  Our study thus provides an alternative description of the measured $p_T$ dependence of the ratio $\Lambda_{c}^+/D^0$ at RHIC without the inclusion of a large number of unknown charmed baryon resonances as assumed in Ref.~\cite{He:2019vgs}. We have, however, neglected in the present study the space-momentum correlations of both light and charm quarks, which are shown in Ref.~\cite{He:2019vgs} to  help shift the peak of the $\Lambda_{c}^+/D^0$ ratio to higher transverse momentum. Also, the present study is based on a blast-wave model for light quarks.  It is thus important to verify the validity of the results and conclusions from the present study by using the phase-space distributions of light and charm quarks from more realistic models.   Since the  light or strange baryon to meson ratio at $p_T^{} \approx 4-7$~GeV in the quark coalescence approach without the flow effect is very small compared to the experimentally measured value~\cite{Minissale:2015zwa}, it will also be very interesting to check if the inclusion of  the flow effect  can help resolve this discrepancy.

{\it Note added in proof.} After the completion of present study, similar results and conclusions have been obtained from a further improved model that uses the quark phase-space distributions from a realistic dynamic model and also includes the production of $p$-wave charmed hadron resonances from quark coalescence~\cite{Cao:2019iqs}.

%\appendix

\begin{acknowledgments}
We are grateful to Xin Dong and Shuai Y.F. Liu for helpful discussions. This work was supported in part by the U.S. Department of Energy under Contract No. DE-SC0015266, the Welch Foundation under Grant No. A-1358, and the National Research Foundation of Korea (NRF) under Grants No.~NRF-2018R1D1A1B07048183, No.~NRF-2018R1A6A1A06024970, No.~NRF-2018R1A5A1025563, and No.~NRF-2019R1A2C1087107.
\end{acknowledgments}

\end{document}